\documentclass{elsart}

\usepackage{graphicx,epsfig}
\usepackage{epsfig}
\input epsf
\usepackage{amssymb}

\newcommand{\be}{\begin{equation}}
\newcommand{\ee}{\end{equation}}
\newcommand{\vep}{\mbox{\boldmath${\rm p}$}}

\newcommand{\veq}{\mbox{\boldmath${\rm q}$}}

\begin{document}
\hspace{9.8 cm}FZJ--IKP(TH)--2003--13
\begin{frontmatter}

  \title{\bf Evidence that the $a_0(980)$ and $f_0(980)$ are not elementary
    particles} \author{V. Baru$^{1,2}$, J. Haidenbauer$^2$, C. Hanhart$^2$}
  \author{Yu. Kalashnikova$^1$, A. Kudryavtsev$^1$}

\address{$^1$Institute of Theoretical and Experimental Physics,
117259,\\ B.\-Cheremushkinskaya 25, Moscow, Russia, \\
$^2$Institut f\"ur Kernphysik, Forschungszentrum
J\"ulich, D-52425 J\"ulich, Germany}

\begin{abstract}
  We study the interesting problem of whether it is possible to distinguish
  composite from elementary particles. In particular we generalize a
  model-independent approach of S. Weinberg to the case of unstable particles.
  This allows us to apply our formalism to the case of the $a_0(980)$ and
  $f_0(980)$ resonances and to address the question whether these particles
  are predominantly genuine, confined quark states (of $\bar q q$ or $qq\bar
  q\bar q$ structure) or governed by mesonic components.
\end{abstract}
\end{frontmatter}

\section{Introduction}

In the mid-sixties S. Weinberg suggested an elegant way to decide whether 
a given particle is composite or elementary \cite{SWein}. The idea was 
applied to the case of the deuteron, and it was shown that the physical 
deuteron is not an elementary particle. In order to prove it the 
field renormalization factor $Z$, $0 \le Z \le 1$, was used,
which is the probability of finding the physical deuteron $|d \rangle$
in a bare elementary-particle state $|d_0 \rangle$, $Z= |\langle d_0|d 
\rangle|^2$.

If the deuteron is purely elementary, then $Z=1$. On the contrary,     
for a purely composite particle made of a proton and a neutron, $Z=0$. The way to
determine the value of $Z$ from hadronic observables is to express the
scattering length, $a$, and the effective range, $r_e$, in terms of $Z$ as
\cite{SWein}
\be
a=\frac{2(1-Z)}{2-Z}R +O(1/\beta),~~r_e=-\frac{Z}{1-Z}R +O(1/\beta),
\label{wlr}
\ee
where $R=1/\sqrt{m_N\epsilon}$, $\epsilon$ is the deuteron binding energy 
and $1/\beta$ is 
the range of forces. The relations (\ref{wlr}) 
are valid in case of a loosely bound state with small binding energy, so 
that $R \gg 1/\beta$, and they are model-independent in this limit.

If the deuteron is composite, then $Z=0$ , $a=R$, $r_e \approx O(m_{\pi}^
{-1})$, with $r_e > 0$. The limit $Z\to 0$ is in
agreement with the experimental values for $a$ and $r_e$: $a$=+5.41 fm,
$r_e$=+1.75 fm. It means that the deuteron is indeed mostly a composite system
made of a proton and a neutron \cite{SWein}. If the deuteron had considerable
admixture of elementary-particle component, then $r_e$, in accordance with
Eqs. (\ref{wlr}) would be large and negative!

Three requirements are needed for the analysis to 
be applicable \cite{SWein}: i) the particle must couple to a two-body
channel with threshold close to the nominal mass; 
ii) this two-body channel must have zero orbital momentum; iii) the 
particle must be stable, otherwise the
analysis cannot be performed in terms of a real $Z$, and the probabilistic 
interpretation is lost. At present one can find particles for which (i) 
and (ii) are satisfied but, usually, (iii) is not satisfied, and one 
tends to agree with S. Weinberg that ``nature is doing her best to keep us 
from learning whether the elementary particles deserve that title''.

In the present paper we adapt Weinberg's idea for the case of unstable
particles. In particular, we consider the spectral density $w(E)$ which is
the probability for finding a bare elementary state in the continuum 
introduced in \cite{bhm}. We show that the integral of this 
quantity over the resonance region serves as a natural generalization 
of Weinberg's variable $Z$. 
We also consider the near-threshold singularities
of the scattering amplitude. The concept of ``pole counting'' as a tool for
resonance classification was formulated by D. Morgan in \cite{morgan}. 
We demonstrate that the "pole counting" scheme is directly related to
Weinberg's analysis in terms of $Z$ (or its continuum counterpart 
$w(E)$). Thus the notion of pole counting can be put on a much more 
quantitative basis. 

The most obvious case to apply our approach to are the $a_0(980)$/$f_0(980)$ 
resonances, the most controversial objects of meson spectroscopy.
Quark models \cite{Isgur} predict $1^3P_0$ $q \bar q$ states made 
of light quarks to exist at about 1 GeV (see also recent developments in
\cite{Alla}), and the $f_0(980)$ and $a_0(980)$ are natural candidates for 
such states. However, as these states are rather close to the $K \bar K$ 
threshold, a significant $qq \bar q \bar q$ affinity is expected 
from a phenomenological point of view \cite{CT}, either in the form of 
compact $qq \bar q \bar q$ states \cite{Achasov} or in the form of 
loosely bound $K \bar K$ states.

A central question of the ongoing debate regarding the nature of 
those scalar resonances is, whether there are sufficiently strong
$t$--channel forces so that $K\bar K$ molecules are formed, 
as advocated in Refs. \cite{WeiIs,Jue,Markushin}, or whether the
meson--meson interaction is dominated by s--channel states. In the
former case the $f_0(980)$ and $a_0(980)$ would be composite 
particles whereas in the latter case they would be elementary 
states linked to quark-gluon dynamics. 
(For a compact presentation of the
discussion we refer to Ref. \cite{IsSp}.) To complicate matters further,
there is no consensus on the manifestation of the $s$-channel state. 
There are claims that this $s$--channel state (of $q \bar q$ 
\cite{Alla,anis} or of four quark nature \cite{Achasov}) manifests itself 
as genuine confined quark state, while in 
Refs. \cite{port,Tornqvist,BiPe} it is found
that a strong coupling to the $K \bar K$ channel causes 
unitarity 
effects which lead to large hadronic component in the wave function of 
the $a_0/f_0$ mesons. More recent
works stress the relevance of chiral symmetry for the
interactions of the lightest pseudo--scalars as well as for the 
formation of corresponding bound states \cite{Oller}.
Up to date the
observed features of the $a_0$ and $f_0$ states could be explained by both the
existence of bare confined states strongly coupled to mesonic channels and/or
by potential-type interactions. Thus the question arises whether it is
possible to distinguish, at a quantitative level, between different
assignements for $a_0$/$f_0$. In the present paper we show, that 
this could be indeed the case.
 
The decays $f_0 \rightarrow \pi\pi$ and $a_0 \rightarrow \pi\eta$ are known 
to be the main source of the width for these mesons. The results of data
analyses are often presented in terms of the Flatt{\`e} parameters 
\cite{Fl}. We argue that the Flatt{\`e} parametrization 
conveniently offers the possibility to connect the singularities of the
scattering amplitude with the manifestations of bare states and to calculate
the spectral density and admixture of bare states in the near-threshold 
region. However, we also show the limitations of the Flatt{\`e} parametrization
and point to situations where its use is doomed to fail. 

The paper is structured in the following way: In the
next section we outline our formalism. In section 3
we discuss the connection of the field renormalization
factor $Z$ and the spectral density $w(E)$ with the 
pole counting scheme proposed by D. Morgan and also with
the effective range parameters. 
In section 4 we apply our formalism to the case of the
$a_0(980)$ and $f_0(980)$ resonances. 
The paper ends with a short summary.

\section{\bf Dynamics of coupled channels}

First we briefly review the dynamics in a coupled channel system,
and show that the small binding limit of it
yields, in the elastic case, the relations (\ref{wlr}). 
It is assumed that the hadronic state is represented 
symbollically as 
\be
|\Psi \rangle = \left(\sum_{\alpha}c_{\alpha}|\psi _{\alpha}\rangle\atop \sum_i
\chi_i |M_1(i)M_2(i) \rangle \right),
\label{state}
\ee
where the index $\alpha$ labels bare confined states  
$|\psi_{\alpha}\rangle$ with the probability amplitude $c_{\alpha}$, and 
$\chi_i$ is the wave function in 
the $i$-th two-meson channel $|M_1(i)M_2(i)\rangle$.  
In coupled-channel models like in Refs. \cite{port,Tornqvist,Markushin}
these bare states are taken to be of  $q \bar q$ nature, but, 
for example, totally confined $qq \bar q \bar q$ states
\cite{Achasov} may be considered as well. One should only ensure the 
orthogonality 
condition $\langle \psi_{\alpha}|M_1(i)M_2(i) \rangle=0$ to be fulfilled.  
The wave function $|\Psi\rangle$ obeys the equation 
\be
\hat {\mathcal H} |\Psi\rangle = E |\Psi\rangle,~~
\hat {\mathcal H} = \left(\begin{array}{cc}
\hat{H}_c&\hat{V}\\
\hat{V}&\hat{H}_{MM}
\end{array}
\right),
\label{ham}
\ee
where $\hat {H}_c$ defines the discrete spectrum of bare states, 
$\hat {H}_c |\psi_{\alpha}\rangle = E_{\alpha} |\psi_{\alpha}\rangle$,
$\hat{H}_{MM}$ includes the free-meson part as well as direct 
meson-meson interaction (e.g., due to $t$- or $u$-channel exchange forces), 
and the term $\hat {V}$ is responsible for dressing the bare states.

Let us start from the simple two-channel case of the 
coupled-channel equation (\ref{ham}). 
Namely, we consider only one bare state $|\psi_0 \rangle$
and only one hadronic channel ($|K \bar K \rangle $). In addition, we assume
proper field redefinitions to be performed, such that the residual $\bar KK$
interaction can be treated perturbatively \cite{SWein}.
The interaction part is specified by the transition form factor 
$f_K(\vep)$,
\be
\langle \psi_0|\hat {V}|K \bar K \rangle = f_K(\vep),
\label{ff}
\ee 
where $\vep$ is the relative momentum in the mesonic system.  For the
mesonic channel in the relative $S$-wave the form factor depends on the
modulus of $p$.  We require the function $f_K$ to decrease with $p$, with some
range $\beta$ whose scale is set by the range of forces or---speaking in quark
language--by the internal size scales of the quark wave functions.
Both scenarios lead to the estimate for $\beta$ to be of the order of a few
hundred MeV. 
One immediately arrives at a system of coupled equations for
$c_0(E)$ and $\chi_E(\vep)$: 
\be 
\left\{
\begin{array}{c}
c_0(E)E_0 + \int f_K(p)\chi_E(\vep)d^3p = c_0(E)E,\\
{}\\
\frac{p^2}{m}\chi_E(\vep) + c_0(E)f_K(p) = E \chi_E(\vep).
\end{array} 
\right.
\label{system}
\ee 
Here $m$ is the meson mass and $2m+E_0$ is the mass of the bare state. 
Throughout the paper we take $m=(m_{K^+}+m_{K^0})/2=495.7$ MeV. In what 
follows we will be interested in the near-threshold phenomena in the 
$S$-wave $K \bar K$ system, so that nonrelativistic kinematics employed in 
Eq. (\ref{system}) is justified.
The system of Eqs. (\ref{system}) can be easily solved yielding for the $K 
\bar K$ scattering amplitude the form
\be
F_{K \bar 
K}(k,k;E)=-\frac{2\pi^2mf_K^2(k)}{E-E_0+g_K(E)},~~k=\sqrt{mE},~~
\frac{d\sigma}{d\Omega}=|F_{K \bar K}|^2,
\label{amp}
\ee 
where
\be 
g_K(E)=\int \frac{f_K^2(p)}{\frac{p^2}{m}-E-i0} d^3p.
\label{g}
\ee
Let the system possess a bound state with the energy $-\epsilon$, 
$\epsilon
> 0$. The binding energy $\epsilon$ fulfills the 
equation
\be
-\epsilon -E_0 + g_K(-\epsilon)=0. 
\label{E_0}
\ee
The
wave function $|\Psi_B\rangle$ of this bound state takes the form
\be
|\Psi_B\rangle=\left(\cos\theta|\psi_0\rangle\atop 
\sin\theta\chi_B(\vep)|K \bar K\rangle
 \right),~ \langle\Psi_B|\Psi_B\rangle=1,~~\cos\theta=<\psi_0|\Psi_B>,
\label{twf}
\ee where $\chi_B(\vep)$ is normalized to unity, and $\cos\theta$ defines the
admixture of the bare elementary state in the physical state $|\Psi_B\rangle$.
Clearly, $\cos^2\theta$ equals the field renormalization $Z$ already discussed
in the beginning. The angle $\theta$ is given by \be \tan^2\theta=\int
\frac{f_K^2(p)d^3p}{(\frac{p^2}{m}+\epsilon)^2}.
\label{angle}
\ee
Let us now demonstrate that in the small binding limit $\sqrt{m\epsilon} 
\ll \beta$ it is possible to express the effective range 
parameters in terms of the binding energy $\epsilon$ and angle $\theta$ 
in a model-independent way. In the region $\sqrt{m|E|} 
\ll \beta$ the integral $g_K(E)$ can be written as
\be
g_K(E)=\bar E_K
+2\pi^2imf_{K0}^2\sqrt{mE}
+O(m^2E/\beta^2),\\
\label{gs}
\ee
where \ $\displaystyle \bar E_K = 4\pi m\int_0^\infty f_K^2(p)dp$
\ and $f_{K0}=f_K(0)$. The integral $\bar E_K$ depends on the explicit form of
the transition form factor and its actual value is thus renormalization scheme
dependent.  However, only the difference $E_0-\bar E_K$ enters the expressions
for observables and the scheme dependence of $\bar E_K$ is absorbed in $E_0$.
In the small binding limit the expression Eq.  (\ref{angle}) takes the form
\ \ $\displaystyle \tan^2\theta = \pi^2m^2f_{K0}^2/\sqrt{m\epsilon}.$ \ \ 
Now 
one easily reads from
Eq. (\ref{amp}) the result given in Eq. (\ref{wlr}), and these expressions 
indeed do not depend on the explicit form of the form factor $f$.
 
The factor $Z$ is the cornerstone of the analysis \cite{SWein}. 
Obviously, the same information is also contained in its continuum counterpart $w(E)$, 
the spectral density of the bare state \cite {bhm}, given by the 
expression
\be
w(E)=2\pi m k |c_0(E)|^2,
\label{w}
\ee
where $c_0(E)$ is found from the system of Eqs. (\ref{system}),
\be
c_0(E) = \frac{f_K(k)}{E-E_0+g_K(E)},
\label{ce}
\ee
and which is definied for energies $E > 0$.
$w(E)$ defines the probability to find the bare state in the 
continuum wave function $|\Psi_E \rangle$ and, as we will see, can be easily
generalized to a situation where inelastic channels are present.
 The 
normalization condition for the distribution $w(E)$ 
follows from the completeness 
relation for the total wave function (\ref{state}) projected onto bare 
state channel. It reads 
\be
\int_0^{\infty}w(E)dE=1-Z~~ or ~~1,
\label{intw}
\ee
depending on whether there is a bound state or not.
Rewriting Eq. (\ref{w}) as 
\be
w(E)=\frac{1}{2\pi i} \left (
\frac{1}{E-E_0+g^{*}_K(E)}-\frac{1}{E-E_0+g_K(E)} \right)
\label{mila}
\ee
the integral (\ref{intw}) can be easily
calculated (for details 
see Ref. \cite{bhm}).
One immediately sees that, for the case of a bound state,  
all the information on the factor $Z$ is encoded, due to 
Eq. (\ref{intw}), in $w(E)$ too.
 On the other hand, 
$w(E)$ can be applied for the case of resonance states as well.

The generalization of the formulae above to the multichannel case is
straightforward. To this end one should
introduce, in addition to (\ref{ff}), a transition form factor 
$f_P(\veq)$ which couples the bare state to light
pseudo--scalars ($\pi\pi$ or $\pi\eta$) with relative momentum $\veq$. 
The
$K \bar K$ scattering amplitude is then
given again by Eq. (\ref{amp}) with the obvious replacement
\be
g_K(E) \rightarrow g_K(E)+g_P(E), ~~g_P(E)=-\int
\frac{f_P^2(p)}{E_{tot}-E_P(p)+i0}d^3p,
\label{gp}
\ee 
where $E_P(p)$ denotes the (relativistic) energy of the two light
pseudo--scalars in the intermediate state in the center of mass system and
$E_{tot}=E_P(q)=E+2m$. The same 
replacement should be done in Eq. (\ref{mila}) for
the spectral density.  Thus, the spectral density is now defined below as well
as above the $K \bar K$ threshold, and, if the exotic possibility of the
existence of $\pi\pi$ or $\pi\eta$ bound states is ignored, is normalized to
unity with a lower limit of integration in Eq. (\ref{intw}) that corresponds
to the threshold of the light pseudo--scalar channel.  It is assumed here that
there is no direct interaction in all mesonic channels.

Since we are interested only in the phenomena near the $K \bar K$ threshold
one can introduce some simplifications. Specifically, one can make use of the
smooth dependence of the integral $g_P(q)$ on the momentum $q$ and replace
$g_P(q)$ by $\bar E_P +\frac{i}{2} \Gamma_P$. The quantities $\bar E_P$ and
$\frac{1}{2} \Gamma_P$ are the real and imaginary parts of $g_P(\bar q)$,
where $\bar q$ is the momentum in the light pseudo--scalar channel averaged
over the $K \bar K$ near-threshold region. This simplification allows also to
avoid problems arising from the necessity of a relativistic treatment of the
light pseudo--scalar channel. We come back to these corrections later in the
manuscript.

In the near-threshold region the expression for the $K \bar K$ scattering
amplitude with inelasticity can be
written as
\be
F_{K \bar K}= -\frac{1}{2k} \frac{\Gamma_K}{E-E_f+i\frac{\Gamma_K}
{2}+i\frac{\Gamma_P}{2} + O(m^2E/\beta^2)},
\label{ampin}
\ee 
where
$$
E_f=E_0-\bar E_K -\bar E_P,~~\Gamma_K
=\bar g_{K \bar K}\sqrt{mE},~~\bar g_{K \bar K}=4\pi^2mf_{K0}^2.
$$
If the terms $O(m^2E/\beta^2)$ in Eq. (\ref{ampin}) are omitted, one immediately 
recognizes in this form the Flatt{\`e} parametrization \cite {Fl} of the 
near-threshold $K \bar K$ amplitude.\footnote{Our dimensionless 
coupling constant $\bar g_{K \bar K}$ is related to the dimensional  
coupling constant $g_{K \bar K}$ commonly used in the literature as
$\bar g_{K \bar K}=\frac{g^2_{K \bar K}}{8\pi M^2_R}$, where $M_R$ is the 
mass of the resonance.} It is convenient to define the mass of the 
resonance $M_R$ as the mass at which the real part of the 
denominator of Eq. (\ref{ampin}) is zero:
\be
M_R=2m+E_R,~~E_R-E_f-\frac{1}{2}\bar g_{K \bar K}
\sqrt{-mE_R}\Theta(-E_R)=0.
\label{mass}
\ee    
In the same way we obtain
the near-threshold expression for the spectral density:
\be
w(E)=\frac{1}{2\pi}\frac{\Gamma_P+
\bar g_{K \bar K}\sqrt{mE}\Theta(E)}
{(E-E_f-\frac{1}{2}\bar g_{K \bar
K}\sqrt{-mE}\Theta(-E))^2+
\frac{1}{4}(\Gamma_P+\bar g_{K \bar K}\sqrt{mE}\Theta(E))^2}.
\label{wfl}
\ee
Eq. (\ref{wfl}) expresses the spectral density $w(E)$ in terms 
of hadronic observables (Flatt{\`e} parameters), 
just in the same way as Weinberg's 
factor $Z$ is expressed in terms of hadronic observables (effective range 
parameters) via Eqs. (\ref{wlr}). Thus, Eq. 
(\ref{wfl}) generalizes Weinberg's result to the case of 
unstable particles.

Note that 
formula (\ref{ampin}) can be easily rewritten as effective 
range 
expansion, and the scattering length $a$ and the effective range $r_e$ are
then given in terms of the Flatt{\`e} parameters by
\be
a=-\frac{\bar g_{K \bar K}}{2(E_f-i\frac{\Gamma_P}{2})}, ~~
r_e=-\frac{4}{m\bar g_{K \bar K}}.
\label{ef}
\ee
Therefore, in the Flatt{\'e} approximation with inelasticity the scattering 
length becomes complex, and $r_e$ remains real and is negative.
In this context it is worth mentioning 
that $r_e$ is also negative in Weinberg's case, once the 
corrections of order $O(1/\beta)$ are omitted in Eq. (\ref{wlr}).

\section{\bf Spectral density and pole counting}

The singularities of the $K \bar K$ scattering
amplitude (Eq. (\ref{ampin})) are given by the zeros of the denominator. 
The pole positions 
can be expressed in terms of the effective range parameters (\ref{ef}): 
\be
k_{1,2}=\frac{i}{r_e} \pm \sqrt{-\frac{1}{r_e^2}+\frac{2}{ar_e}}
\label{roots}
\ee

In the purely elastic case and in the presence of a bound state, one 
has $E_f < 0$ and therefore $a > 0$, cf. Eq. (\ref{ef}). 
The poles are then located at the imaginary axis
of the $k$-plane, one in the upper and the other one in the lower half plane.
In terms of Weinberg's variable $Z$ the pole positions are given by
\be
k_1=i\sqrt{m\epsilon}, ~~k_2=-i\sqrt{m\epsilon}\frac{2-Z}{Z}.
\label{wroots}
\ee
The first pole in (\ref{wroots}) is located in the near-threshold 
region. (Recall that we consider the small binding energy limit.)  
For a deuteron-like situation, i.e. 
for $Z \ll 1$, the second pole is far from the threshold and even moves to 
infinity in the limit $Z \rightarrow 0$. 
On the other hand, if $Z$ is
close to one, i.e. if there is considerable admixture of an elementary state in the 
wave function of the bound state, both poles are near threshold. In the 
limiting case $Z \rightarrow 1$ (pure bare state) the poles are located equidistant 
from the point $k=0$. Obviously there is a one-to-one correspondence between 
$Z$ and the ``pole counting'' arguments of \cite{morgan}:
a bound state with large admixture of a bare state (large $Z$) manifests 
itself 
as two near-threshold pole singularities, while a deuteron-like
state corresponds to a small $Z$ and gives rise to only one nearby pole.

Let us now consider the case $E_f > 0$ and go over to the spectral density $w(E)$. 
Since the singularities of $c_0(E)$ (\ref{ce}) coincide with the ones 
of the amplitude, the behaviour of $w(E)$ is also governed 
by the poles (\ref{roots}). Specifically, 
one expects the spectral density to be enhanced in the vicinity of poles. 
Thus, if both poles (\ref{roots}) are located in the 
near-threshold region, the spectral density in this region would be large 
-- and indicates thereby that the bare state admixture in the 
near-threshold resonance is large. 
If, on the contrary, there is only one near-threshold 
pole, a considerable part of the spectral density is smeared over a much 
wider energy interval, which is a signal that the 
bare state admixture in the near-threshold resonance is small.

As pointed out in the Introduction, in case of a weakly-bound
state there is also a unique relation between $Z$ and the effective
range parameters. Specifically, $r_e$ is large and negative if there
is a large admixture of the bare state in the physical bound state
wave function. It is interesting to see what happens in 
the case of a near-threshold resonance state, i.e. when $E_f > 0$.
First we see from Eq. (\ref{ef}) that then the scattering length 
$a$ is negative. 
For small values of $|r_e|$ which fulfil the relation 
$|a| > 2|r_e|$ both poles are located on the negative imaginary axis,
cf. Eq. (\ref{roots}). 
When $r_e$ is larger so that $|a| < 2|r_e|$ and also 
negative then the pole positions acquire a real part. 
For very large and negative $r_e$,  
$|a| \ll 2|r_e|$, the real part of both poles positions is much 
larger than the imaginary one -- which corresponds to the case of a 
well-pronounced narrow resonance at the energy $E_f$. 
Thus, a large and negative $r_e$ corresponds to a  
spectral density that is strongly enhanced around the resonance region 
and, consequently, to the case of a large admixture of the bare state. 
On the other hand, for small (positive or negative) $r_e$ there is only one 
near-threshold pole, 
the spectral density is not strongly enhanced in the near-threshold 
region and, accordingly, the resonance contains a large mesonic component.

The inclusion of an inelastic channel (in our case $\pi\pi$ or $\pi\eta$)
does not change the picture qualitatively. Specifically, if 
the poles of the $K\bar K$ amplitude are again close to each other and to 
the threshold then the effective range should be negative and large.
This will be directly reflected in the 
Flatt\`e parameter $\bar g_{K \bar K}$, which then should be 
small (see Eq. (\ref{ef})).

\section{\bf Application to the $a_0$ and $f_0$ mesons}

Let us now apply our formalism to the $a_0$ and $f_0$ mesons. 
We start from some recently published Flatt{\`e} 
and Flatt{\`e}-type representations of the corresponding $\pi\eta$ and 
$\pi\pi$ spectra. For convenience we have summarized the employed
Flatt{\`e} parameters of the $a_0$ and $f_0$ mesons 
(together with the references from where the values are taken) 
in the first four columns of Tables 1 and 2. 

After the discussion in the previous section
it should be clear that it is rather instructive to study the
behaviour of $w(E)$ over the region of interest, i.e. over the energy
region containing the $K\bar K$ threshold. Thus, it is 
useful to introduce an integrated quantity by
\be
W_{a_0 (f_0)}=\int^{50 MeV}_{-50 MeV} w_{a_0 (f_0)}(E)dE \ .
\label{W}
\ee
Obviously, $W_{a_0 (f_0)}$ is the probability for
finding the bare state in the specified energy interval -- which we have chosen to
be roughly twice as large as the peak width of the $a_0$ and $f_0$ mesons. 
Recall that $w(E)$ is normalized in such a way that the integral over the
whole energy range amounts to unity.  For a well-pronounced ``pure'' resonance
of Breit-Wigner type, i.e. with a negligibly small coupling $\bar g_{K \bar
  K}$ in Eq. (\ref{wfl}), one gets for the integral $W$ a value of
$\frac{2}{\pi}\arctan2 \approx 0.70$ for the considered energy interval.
Therefore, the deviation from this value provides a direct measure for the
admixture of mesonic components in the $a_0/f_0$ mesons.

The results for the $a_0$ meson -- 
the positions of the poles $k_{1,2}$ in the complex plane $k$,
$W_{a_0}$, and the $K\bar K$ scattering lengths and
effective ranges calculated via Eq. (\ref{ef}) -- 
are given in the last six columns of Table 1, 
while those for the $f_0$ meson can be found in Table 2.

\begin{table}[t]
\begin{center}
\caption{Parameters and results for the $a_0$ meson. 
The values $M_R$, $\Gamma_{\pi\eta}$ and $E_f$ are
given in MeV, $r_e$ and $a$ in fm, and $k_1$ and $k_2$ in MeV/c.}
\vskip 0.2cm
\begin{tabular}{|c|c|c|c||c|c|c|c|c|c|}
\hline
Ref.&$M_R$&$\Gamma_{\pi\eta}$&$\bar g_{K\bar
K}$&$E_f$&$r_e$&$a$&$k_1$&$k_2$&$W_{a_0}$\\
\hline
\cite{Teige}&1001&70&0.224&9.6&-7.1&-0.16-i0.59&-104+i55&104-i111&0.49\\
\cite{Bugg}&999&146&0.516&7.6&-3.1&-0.07-i0.69&-134+i71&134-i199&0.29\\
\cite{AchKi}&1003&153&0.834&11.6&-1.9&-0.16-i1.05&-129+i44&129-i250&0.24\\
\cite{AchKi}&992&145.3&0.56&0.6&-2.8&-0.01-i0.76&-126+i73&126-i212&0.29\\
\cite{KLOE}&984.8&121.5&0.41&-18.0&-3.
9&0.18-i0.61&
-102+i97&102-i199&0.36\\
\hline
\end{tabular}
\label{Table 1}
\end{center}
\end{table}

\begin{table}[t]
\begin{center}
\caption{Parameters and results for the $f_0$ meson.
The values $M_R$, $\Gamma_{\pi\pi}$ and $E_f$ are
given in MeV, $r_e$ and $a$ in fm, and $k_1$ and $k_2$ in MeV/c.}
\vskip 0.2cm
\begin{tabular}{|c|c|c|c||c|c|c|c|c|c|}
\hline
Ref.&$M_R$&$\Gamma_{\pi\pi}$&$\bar g_{K\bar K}$&$E_f$&$r_e$&
$a$&$k_1$&$k_2$&$W_{f_0}$\\
\hline
\cite{SND}&969.8&196&2.51&-151.5&-0.63&1.15-i0.74&-58+i107&58-i729&0.17\\
\cite{CMD2}&975&149&1.51&-84.3&-1.05&0.99-i0.88&-65+i97&65-i477&0.23\\
\cite{KLOE}&973&253&2.84&-154&-0.56&1.09-i0.89&-69+i100&69-i804&0.14\\
\cite{AchGu}&996&128.8&1.31&+4.6&-1.22&-0.14-i 1.99&-84+i17&84-i351&0.21\\
\hline
\end{tabular}
\label{Table 2}
\end{center}
\end{table}

First, let us point out 
that the couplings $\bar g_{K \bar K}$ for the $a_0$ and $f_0$ mesons
differ drastically. In case of the $a_0$ it is, in average, significantly 
smaller than for the $f_0$. 
Accordingly, the effective range $r_e$ is fairly large for the $a_0$ 
case and much smaller for the $f_0$ case, as can be seen from the Tables.

The pole positions in the complex $k$ plane, cf. Tables 1 and 2,
are shown graphically in Fig. 1. For the $a_0$ meson the poles
look roughly equidistant, and both poles are still close to the
real axis. Thus, it is expected that both singularities influence the 
behaviour of the near-threshold amplitude. For the
$f_0$ meson there is only one pole close to the physical region.

Let us now examine the spectral densities shown in Fig. 2. 
The $a_0$ meson appears as an above-threshold phenomenon
(with the exception of the fit of Ref. \cite{KLOE}), with a spectral
density peaked at the $K \bar K$ threshold. 
The density $w_{a_0}(E)$ for various Flatt{\`e} representations look rather similar, 
with exception of the
curve for the result of Ref. \cite{Teige}. Here the coupling $\bar g_{K \bar K}$
is very small. This fit leads to almost equidistant positions
of the poles, and the near-threshold fraction of $w(E)$ is sizable, as reflected in
the large value of $W_{a_0}$. Still, 
even for this fit $W_{a_0}$ does not exceed 50\%, while the average for all considered
Flatt{\`e} parametrizations is about 30\%. Clearly, this 
means that the $a_0$ (980) should not be considered as a pure 
quark state, but definitely has a sizeable 
admixture of mesonic components.

The $f_0$ meson appears predominantly as a sub-threshold resonance 
(again with one exception \cite{AchGu}), 
with the spectral density being peaked a few MeV below the $K\bar K$ threshold. 
Above the $K\bar K$ threshold $w(E)$ is very small. 
The pole positions are not equidistant at all as can be seen in Fig. 1b. 
Correspondingly, 
the $f_0$ meson resembles a pure $1^3P_0$ $q \bar q$ state  
even less than the $a_0$ meson.

As a side note we want to emphasize that the Flatt{\`e} parametrizations 
leading to a small $|r_e|$ should be treated with 
caution. This can be seen from 
the relation (\ref{ef}) between the Flatt{\`e} parameters and the 
effective range parameters. With $r_e \rightarrow 0$ all the Flatt{\`e} 
parameters tend to infinity. Obviously, should the physical $K\bar K$ 
effective range $r_e$ indeed be small then the 
determination of $E_r$, $\Gamma_{P}$ and $\bar g_{K \bar K}$ from a 
Flatt{\`e} fit becomes unstable. Moreover, it is possible that 
an equal or even better quality fit may be achieved with positive effective 
range. As mentioned above, a Flatt{\`e} analysis automatically implies that
$r_e < 0$ whereas a small but positive value of $r_e$ is typical for 
the deuteron-like situation, i.e.
for a dynamically generated bound state. 
In any case, it would be desirable to use more general parametrizations
of resonance data that also allow for a positive effective range
and hence can also be applied to potential-type resonances. 
One option would be to employ directly an effective
range expansion to fit the data. Another possibility is to include the
relativistic corrections to the loop integral $g_K(E)$ (see \cite{AchGu}) as
well as the corrections $O(m^2E/\beta^2)$ that appear in our formalism
(but were omitted so far).

\section{\bf Summary}

In the present paper we have investigated the interesting problem of whether 
it is possible to distinguish composite from elementary particles. 
In particular 
we have generalized the model-independent approach of S. Weinberg that
is based on the probabilistic interpretation of the field renormalization $Z$  
to the case of unstable particles (resonances). This could be achieved
by introducing a suitably defined spectral density $w(E)$ which allows to 
analyse not only situations where the physical states lie below the 
threshold but also in the continuum. Both $Z$ as well as $w(E)$ provide
a measure for the admixture of the bare state in the physical state. 
Within this formalism it is possible to address the long-standing question 
whether the $a_0(980)$ and $f_0(980)$ mesons are
genuine quark states or whether they  
contain a dominant admixture of mesonic components. 

Furthermore, 
we have shown that there is a one-to-one correspondence between
the value of $Z$ (or the behaviour of $w(E)$) and the 
pole counting scheme suggested by Morgan \cite{morgan}.
Specifically, in case of values of $Z$ close to one 
(or a strongly enhanced spectral density near threshold) there are 
always two nearly equidistant poles located close   
to the threshold. This situation corresponds to a large admixture
of a bare ($q\bar q$ like) state in the physical bound state or
resonance and it  
manifests itself also by a weak coupling to the 
bare state to the $K \bar K$ channel and by a large 
and negative effective range $r_e$. 
In the opposite case where $Z$ is almost zero or where $w(E)$ is
smeared out over the whole energy range there is only one stable 
pole near the threshold. This situation corresponds to the case
of a dynamically generated bound state or resonance and is
characterized by a strong coupling to the $K \bar K$ channel
and by an effective range that is small or even positive. 

The spectral density $w(E)$ of bare states can be constructed
from hadronic ``observables'' like the Flatt{\`e} parameters 
in a similar way the field renormalization factor $Z$ can be 
extracted from the effective range parameters.  
In order to exemplify the potential of the proposed method
we evaluted $w(E)$, for energies near the $K\bar K$ threshold, for
several recently published Flatt{\`e} parametrizations of measured
$\pi\pi$ and $\pi\eta$ spectra.
Thereby we found that the (near-threshold) probability for finding the 
$a_0$ meson in a bare states is only about 25 to 50\%, indicating that
the $a_0$ should have a significant mesonic component.  
As for the $f_0$ meson, the employed Flatt{\`e} parametrizations 
yielded even smaller values for this probability, namely 
of the order of 20\% or less, providing evidence that its
mesonic component should be indeed rather large. 
Therefore, we conclude that a simple $1^3P_0$ $q \bar q$ or four quark  
assignement for the $a_0$(980) should be considered with caution and it is
certainly questionable for the $f_0$(980).

\medskip

Fruitful discussions with S. Krewald, F. Sassen and J. Speth are greatly 
appreciated. Yu.S. K. and A.E. K 
thank the FZJ-IKP theory group for their hospitality and the grant 
02-02-04001/436 RUS/13/652 for financial support. Yu.S.K. is grateful 
for partial support from the grant NSh-1774.2003.2, and A.E.K. thanks the 
grant RFBR 02-02-16465 for partial support.

\bigskip

\begin{figure}[h]
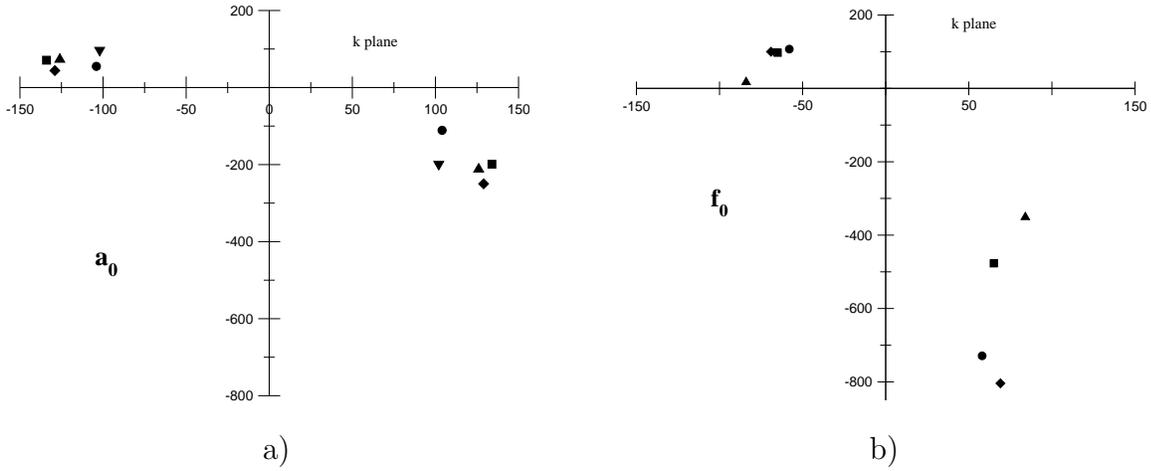

\begin{tabular}{ccc}
\hspace*{-1cm}
\epsfig{file=a0poles2.eps,width=7cm}\hspace*{0.35cm}&&\epsfig{file=f0poles1.eps,width=7cm}\\
\hspace*{-1cm}a) && b)
\end{tabular}
\caption{a) Pole positions for the $a_0$ meson in the complex $k$ plane [in
MeV/c] based on the Flatt{\`e} parameters taken from  
Ref. \cite{Teige} (circles), Ref. \cite{Bugg} (squares), Ref. \cite{AchKi} (diamonds),
Ref. \cite{AchKi} (triangles), and Ref. \cite{KLOE} (reversed triangles).
b) Pole positions for the $f_0$ meson in the complex $k$ plane [in 
MeV/c] based on the Flatt{\`e} parameters taken from 
Ref. \cite{SND} (circles), Ref. \cite{CMD2} (squares), Ref. \cite{KLOE} (diamonds), 
and Ref. \cite{AchGu} (triangles).}
\label{p}
\end{figure}

\begin{figure}[h]
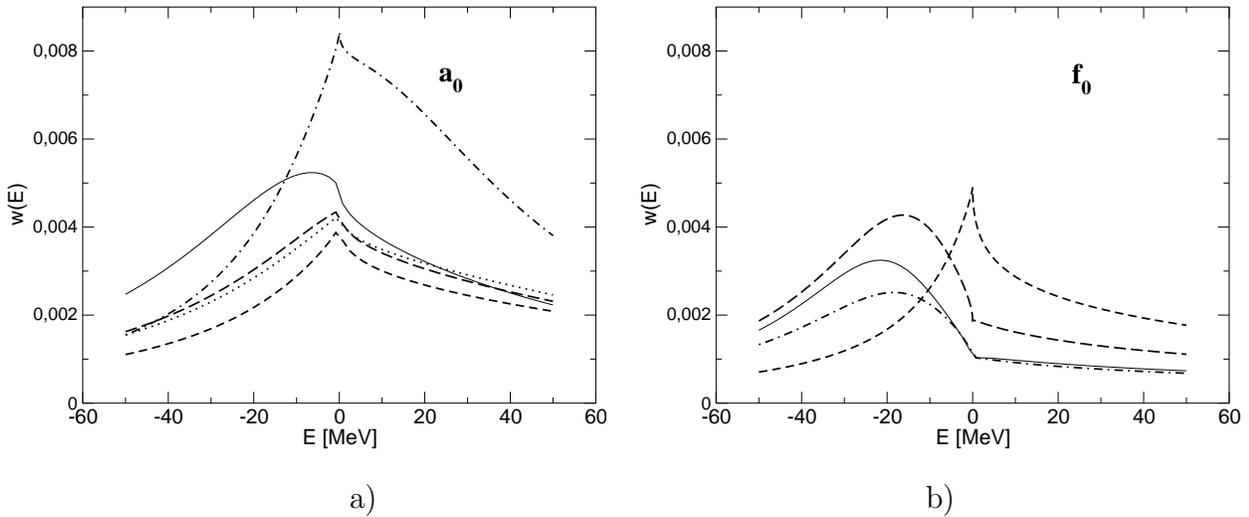

\begin{tabular}{cc}
\hspace*{-1.5cm}\epsfig{file=wa0_1.eps,width=8cm}&\epsfig{file=wf0_1.eps,width=8cm}
\\
a) & b)
\end{tabular}
\caption{a) Spectral densities $w(E)$ for the $a_0$ meson 
based on the Flatt{\`e} parameters taken from  
Ref. \cite{Teige} (dashed-dotted line), Ref. \cite{Bugg} (dotted line), 
Ref. \cite{AchKi} (dashed line), Ref. \cite{AchKi} (long dashed line), and 
Ref. \cite{KLOE} (solid line).
b) Spectral densities $w(E)$ for the $f_0$ meson 
based on the Flatt{\`e} parameters taken from  
Ref. \cite{SND} (solid line), Ref. \cite{CMD2} (long-dashed line), 
Ref. \cite{KLOE} (dashed-dotted line), and Ref. \cite{AchGu} (dotted line).} 
\label{wa}
\end{figure}


\medskip 
  
\begin{thebibliography}{99} 
\bibitem{SWein} S. Weinberg, Phys. Rev. {\bf
130}, 776 (1963); {\bf 131}, 440 (1963); {\bf 137} B672 (1965).
\bibitem{bhm} L.N. Bogdanova, G.M. Hale, and V.E. Markushin, Phys. Rev. C {\bf
44}, 1289 (1991).
\bibitem{morgan} D. Morgan, Nucl. Phys. {\bf A543}, 632 (1992); N. T\"ornqvist,
Phys. Rev. D {\bf 51}, 5312 (1995). 
\bibitem{Isgur} S. Godfrey and N. Isgur, Phys. Rev. D {\bf 32}, 189
(1985). 
\bibitem{Alla} M. Koll, R. Ricken, D. Merten, B. Metsch, and H. Petry,
Eur. Phys. J. A {\bf 9}, 73 (2000); A.M. Badalian, and B.L.G. Bakker, Phys. Rev.
D {\bf 66}, 034025 (2002); A.M. Badalian, hep-ph/0302089.
\bibitem{CT} F.E. Close and N.A. T\"ornqvist, J. Phys. G {\bf 28}, R249
(2002).
\bibitem{Achasov} R. L. Jaffe, Phys. Rev. D {\bf 15}, 267 (1977);
 {\bf 15}, 281 (1977);
 N.N. Achasov, S.A. Devyanin, and G.N. Shestakov, Phys. Lett. {\bf B96}, 168
(1980); M. Alford and  R.L. Jaffe, Nucl. Phys. {\bf B578}, 367 (2000). 
\bibitem{WeiIs} J. Weinstein and N. Isgur,
  Phys. Rev. D {\bf 27}, 588 (1979).
\bibitem{Jue} D. Lohse, J.W. Durso, K. Holinde, and J.Speth, Phys. Lett. 
{\bf B234}, 235 (1990); G. Janssen, B.C. Pearce, K. Holinde, and J.Speth, 
Phys.Rev D {\bf 52}, 2690 (1995).
\bibitem{Markushin} M.P.Locher, V.E.
  Markushin, and H.Q. Zheng, Eur. Phys. J. C {\bf 4}, 317 (1998); V.E.
  Markushin, Eur. Phys. J. A {\bf 8}, 389 (2000).
\bibitem{IsSp} N. Isgur and J. Speth, Phys. Rev. Lett. {\bf 77}, 2332 (1996).
\bibitem{anis} V.V. Anisovich, hep-ph/0208123.
\bibitem{port} E. van Beveren, C. Dullemond, and G. Rupp, Phys. Rev. D {\bf 21}, 772
(1980); G. Rupp, E. van Beveren, and 
M.D. Scadron, Phys. Rev. D {\bf 65}, 078501 (2002); E. van Beveren and G. Rupp,
hep-ph/0304105.
\bibitem{Tornqvist} N.A. T\"ornqvist, Z. Phys. C {\bf 68}, 674 (1995); N.A. 
T\"ornqvist and M. Roos, Phys. Rev. Lett. {\bf 76}, 1575 (1996).
\bibitem{BiPe} M. Boglione and M.R. Pennington, Phys. Rev. D {\bf 65}, 114010 
(2002).
\bibitem{Oller} J.A. Oller and E. Oset, Phys.
  Rev. D {\bf 60}, 074023 (1999).    
\bibitem{Fl} S. Flatt{\`e},
  Phys. Lett. {\bf B63}, 224 (1976).  
\bibitem{Teige} S. Teige et al., Phys.
  Rev. D {\bf 59}, 012001 (2001).  
\bibitem{Bugg} D.V. Bugg, V.V. Anisovich,
  A. Sarantsev, and B.S. Zou, Phys. Rev D {\bf 50}, 4412 (1994).
\bibitem{AchKi} N.N. Achasov and A.N. Kiselev, Phys.
  Rev. D {\bf 68},
014006 (2003).
\bibitem{KLOE}A. Antonelli, hep-ex/0209069.  \bibitem{SND} M.N. Achasov et
  al., Phys. Lett. {\bf B485}, 349 (2000).  \bibitem{CMD2} R.R. Akhmetshin et
  al., Phys. Lett. {\bf B462}, 380 (1999).  \bibitem{AchGu} N.N. Achasov and
  V.V. Gubin, Phys. Rev. D {\bf 63}, 094007 (2001).


\end{thebibliography}
\end{document}